\documentclass[acus]{JAC2003}

\usepackage{graphicx}
\usepackage{booktabs}
\usepackage{subfig}
\usepackage{amsmath}
\usepackage{array}
\usepackage{microtype}
\usepackage[colorlinks=true, citecolor=blue, urlcolor=black, linkcolor=black]{hyperref}
\begin{document}
\title{A microwave paraphoton and axion detection experiment with 300 dB electromagnetic shielding at 3 GHz\thanks{Work supported by the Wolfgang-Gentner-Programme of the Bundesministerium f\"ur Bildung und Forschung (BMBF).}}

\author{M. Betz, F. Caspers, CERN, Geneva, Switzerland}

\maketitle

\begin{abstract}
For the microwave equivalent of ``light shining through the wall'' (LSW) experiments, a sensitive microwave detector and very high electromagnetic shielding is required. The screening attenuation between the axion generating cavity and the nearby detection cavity should be greater than 300 dB, in order to improve over presently existing exclusion limits.
To achieve these goals in practice, a ``box in a box" concept was utilized for shielding the detection cavity, while a vector signal analyzer was used as a microwave receiver with a very narrow resolution bandwidth in the order of a few micro-Hz.
This contribution will present the experimental layout and the results to date.
\end{abstract}

\section{Motivation}
\label{lbl:intro}
The axion is a hypothetical elementary particle, which emerged originally from a proposal by Peccei
and Quinn, intended to solve the strong CP problem \cite{src:WISPy} in theoretical physics.
The axion is neutral, only interacts weakly with matter, has a low mass ($\approx 10^{-4} eV/c^2$), spin zero, and a natural decay constant (to 2 photons) in the order of $10^{17}$ years.
The axion belongs to the family of Weakly Interacting Sub-eV Particles (WISP). Another WISP, closely related to the axion is the paraphoton or hidden photon.
The existence of these WISPs could not be confirmed yet and all experimental efforts to date have so far produced only exclusion results. Nevertheless there is strong motivation to advance the experimental ``low energy frontier'' as the axion is the most popular solution for the strong CP-problem. Many WISPs are also excellent candidates for dark matter and explain numerous astrophysical phenomena. 

\section{Experimental setup}
WISPs can be probed in the laboratory by ``Light Shining through the Wall'' (LSW) experiments.
They exploit the very weak coupling to photons, allowing an indirect proof of the otherwise hidden particles without relying on any cosmological assumptions. 
Previous LSW experiments have been carried out with optical laser light at DESY (ALPS), CERN (OSQAR) and Fermilab (GammeV).

\begin{figure}[t]
\begin{center}
	\includegraphics[width=0.49\textwidth]{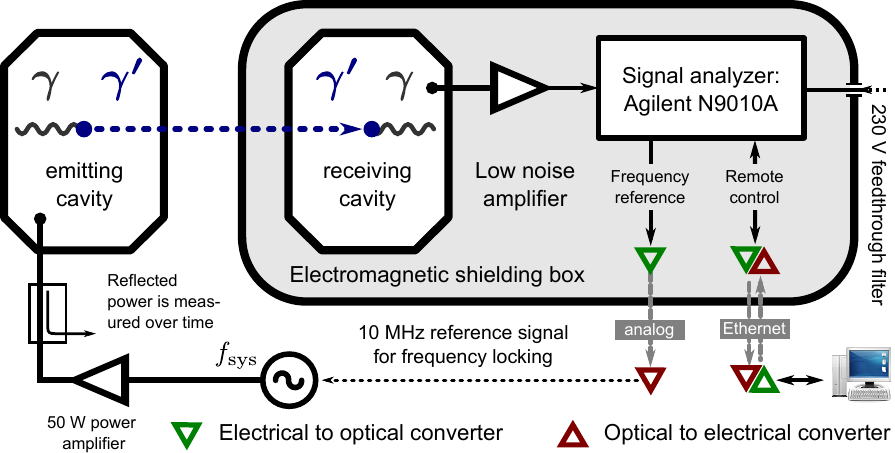}
	\caption{Block diagram of the experiment}
	\label{fig:ovrBlock}
\end{center}
\vspace{-0.75 cm}
\end{figure}

The concept of an optical LSW experiment can be adapted to microwaves \cite{src:hoogUW,src:moiMykonos}. A block diagram of the setup is shown in Fig.~\ref{fig:ovrBlock}, it consists of two identical low loss microwave cavities with a diameter of 140 mm, a height of 120 mm and a spacing between them of 150 mm.
One serves as WISP emitter and is excited by an external microwave source. It develops a strong electromagnetic (EM) field, which corresponds to a large amount of microwave photons $\gamma$. 
Theory predicts that some of these photons convert to paraphotons $\gamma'$ by kinetic mixing (similar to neutrino oscillations) or -- if the cavities are placed in a strong static magnetic field -- to axion-like particles by the Primakoff effect~\cite{src:WISPy}.
Both particles only interact very weakly with matter (similar to neutrinos in this respect) and thereby, in contrast to the photons, can traverse the cavity walls. Some WISPs propagate towards the detection cavity, which is connected to a very sensitive microwave receiver. The reciprocal conversion process transforms WISPs to microwave photons, which can be observed as an excitation of the seemingly empty and well shielded detection cavity. 
Since there is no energy loss associated with the WISP conversion process, the regenerated photons in the detecting cavity have exactly the same energy as the photons in the emitting cavity. Thus, the signal which is coupled out from the detection cavity has the same frequency as the one which is generated on the emitting side, making a narrowband receiving concept feasible.

This paper will focus on the latest exclusion results for \textbf{paraphotons} from the microwave WISP search at CERN. In a future upgrade, an additional magnet will allow the search for axions.

Considering current exclusion limits, it takes $> 10^{24}$ photons on the emitting side to generate one photon on the detection side, making this the most challenging aspect of an LSW experiment.
The expected output power (or photon flux) from the detecting cavity towards the microwave receiver due to paraphotons is given by Eq.~\ref{equ:power},
\begin{align}
	\label{equ:power}
	P_{\mathrm{det}} &= \chi^4 \left(\frac{m_{\gamma'} c^2}{f_{\mathrm{sys}} h}\right)^8 |G|^2 Q_{\mathrm{em}} Q_{\mathrm{det}} P_{\mathrm{em}}
\end{align}
where $Q_{\mathrm{em}}$ and $Q_{\mathrm{det}}$ are the loaded Q factors of emitting and detection cavity, $f_{\mathrm{sys}}$ is the frequency where the experiment is carried out (and to which the cavities are tuned),
$h$ is Planck's constant and $G$ is a dimensionless geometric form factor in the order of 1, describing the position, shape and resonating mode of the cavities \cite{src:JaCaRi}.
The rest mass of hidden photons is a priori unknown and given by $m_{\gamma'}$.
% expressed in $\left[\frac{J}{c^2}\right]$ ($1~\mathrm{J} = 6.24\cdot10^{18}~\mathrm{eV}$). 
The kinetic mixing parameter $\chi$ describes the likeliness of paraphoton - photon oscillations. A previous examination of Coloumb's law indicates that $\chi < 3 \cdot 10^{-8}$ in this energy range.

If there is no significant signal detected, an exclusion result can be produced by
% setting $P_{\mathrm{det}}$ to the minimum amount of microwave power the receiver can still distinguish from noise, then 
determining $\chi$ from the other known values. This provides a convenient way to compare the achieved sensitivity to other experiments.

\begin{figure}[b]
\vspace{-0.5 cm}
\begin{center}
	\includegraphics[width=0.49\textwidth]{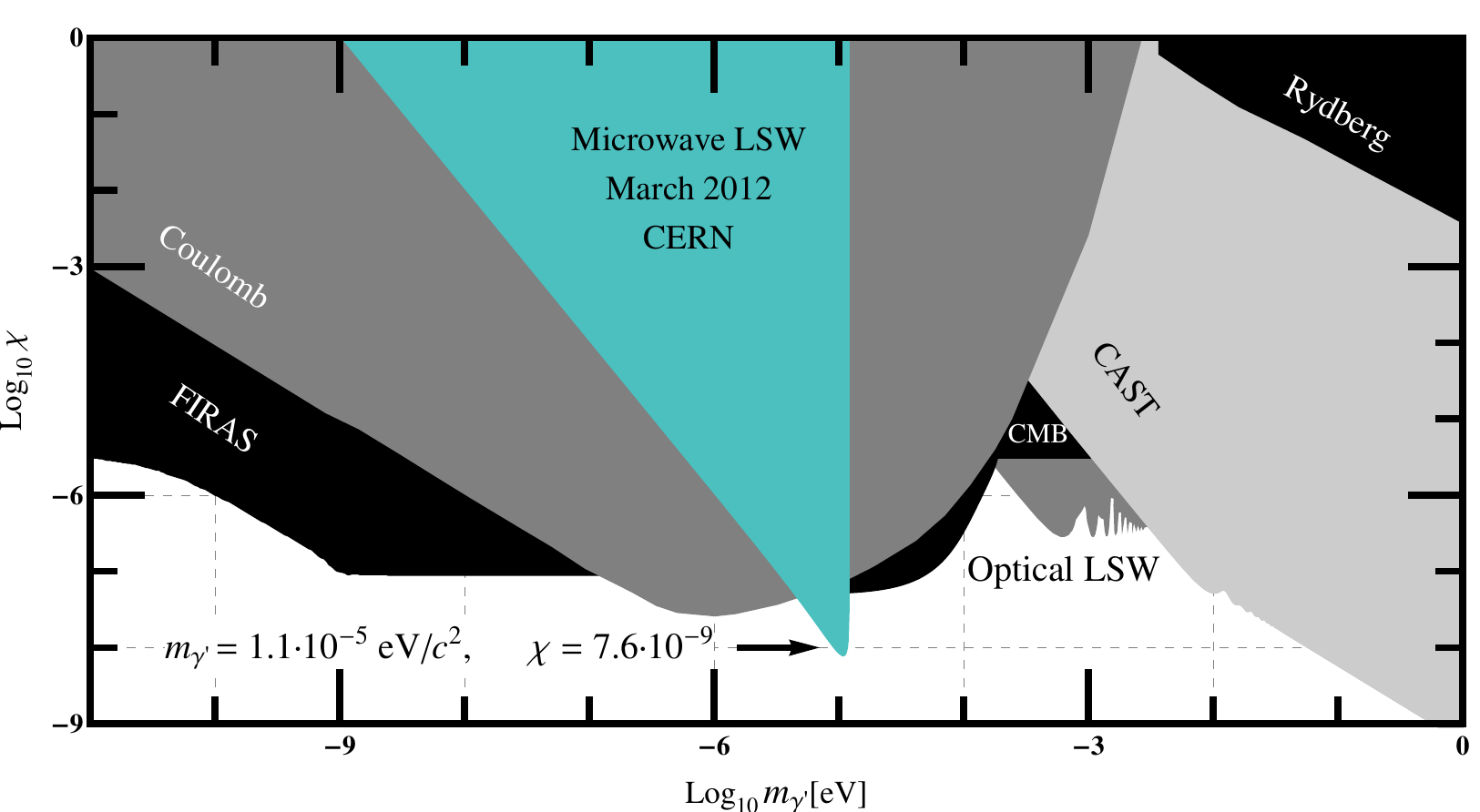}
	\caption{Exclusion limit for paraphotons as a result of the measurement-run at CERN in March 2012, compared to other experiments (details in \cite{src:JaCaRi})}
	\label{fig:exclPlot}
\end{center}
\vspace{-0.5 cm}
\end{figure}

The parameters of the paraphoton experiment as it has been set up and carried out at CERN in March 2012, are summarized in Table~\ref{tbl:param}. As no paraphotons were observed, the corresponding exclusion limit in  comparison to other experiments is shown in Fig.~\ref{fig:exclPlot}.
% The conversion probability can be enhanced by more input power, cavities with higher Q-factors, less distance between them and the choice of a resonating mode which is coupling stronger to paraphotons. Measurements over a longer time capture more reconverted photons, increasing the signal to noise ratio, making them easier to detect and filter out from background noise \footnote{Statistical methods need to be applied as single photon counting has not yet been achieved for microwave photons in the $10^{-5}$ eV energy range}.
\begin{table}[t]
\centering
\caption{parameters of the paraphoton run in March 2012}
\label{tbl:param}
\begin{tabular}{c}
\toprule
$f_{\mathrm{sys}} = 2.9565$ GHz \quad $Q_{\mathrm{det}} = 23620$ \quad $Q_{\mathrm{em}} = 23416$\\[0.1 cm]
$P_{\mathrm{det}} = 8.51 \cdot 10^{-25}$ W \quad $P_{\mathrm{em}} = 37$ W \quad $|G| = 0.222$ \\
\bottomrule
\end{tabular}
\vspace{-0.3 cm}
\end{table}

\section{Engineering aspects}
On the left side of Fig.~\ref{fig:ovrBlock}, a commercial microwave source is shown, which generates a signal on $f_{\mathrm{sys}}$ (see Table~\ref{tbl:param}) which is amplified up to 50 W and drives the emitting cavity on its resonant frequency.

Power is coupled in and out of each cavity with a small inductive coupling loop, adjusted for critical coupling to the TE$_{011}$ mode. This mode has been chosen for its high Q-factor and reasonable coupling to paraphotons compared to other modes. The loaded Q-factor of the silver coated brass cavities has been determined by a network analyzer, their 3 dB bandwidth is $BW_{3 dB} \approx 126$~kHz. A tuning stub perturbing the H-field allows to compensate manufacturing tolerances within a bandwidth of $\approx 10$~MHz.

%\subsection{Electromagnetic shielding}
Shielding is required around the detecting cavity and the microwave receiver to eliminate ambient electromagnetic interference (EMI) and to mitigate coupling to the emitting cavity by simple EM leakage. This would generate false positive results as a signal originating from leakage can not be distinguished from a signal propagating by WISP conversion. Within 15 cm, the field strength must be reduced by at least a factor of $7.7 \cdot 10^{12} = 258 \mathrm{~dB}$ to get meaningful results\footnotemark. \footnotetext{For comparison, the signals from the Voyager 1 space probe, 14 light hours from earth, are currently attenuated by a factor of 240 dB (comparing the field strength in front of sending and receiving antenna)}

The shielding box has been built from a straight piece of WR-2300 waveguide with the inside dimensions 58x29x100~cm. Feeding all RF signals over optical fibres by analog transceivers prevents the propagation of EMI trough ordinary transmission lines. An optical ethernet link is used to remote control the signal analyser.

Shielding effectiveness has been measured by comparing the electric field strength in- and outside the wall with a calibrated electric field probe and a spectrum analyzer. The shielding enclosure provides $\approx 90$~dB and each of the cavities provides an additional $\approx$~110 dB of shielding. The combined EM attenuation is $\approx 310$~dB, making thermal noise the limiting factor for the minimum detectable signal.
 
%\subsection{Tune of the emitting cavity}
For an exclusion result it is necessary to prove the detector is working and actually able to pick up any WISP related signals.
Detection sensitivity will be limited if the resonant frequency of one or both cavities does not equal the system frequency $f_{\mathrm{sys}}$.
This is especially delicate as the cavities are sensitive to temperature variations; their resonant frequency is inversely proportional to the thermal expansion coefficient of their wall material. For brass, a temperature change of $\Delta T = +2$~K leads to a detuning of $\Delta f_{\mathrm{res}} = -112$~kHz and a reflection of around half the input power back towards the amplifier.
% \begin{align}
% \label{equ:therm}
% 	\Delta f_{\mathrm{res}} = - \frac{f_{\mathrm{res}} \cdot \alpha \cdot \Delta T}{1 + \alpha \cdot \Delta T}
% \end{align}

Thermal drift is more critical for the emitting cavity as it has to dissipate up to 50 W of heat by forced air cooling without any external temperature stabilization. Before data taking, the cavity was operated at full power and its resonant frequency was kept constant manually. After around 1 h, the cavity reached thermal equilibrium and no further tuning was necessary. 
% It is interesting to note that thermal equilibrium can only be reached because of a feedback process. An increase in cavity temperature manifest itself in a lower resonant frequency (see Eq.~\ref{equ:therm}) this in turn can lead to more or less power from the amplifier ending up inside the cavity (see Eq.~\ref{equ:cavReflFreq}, depending on the current resonant frequency. If it is lower than $f_{\mathrm{sys}}$, less power is dissipated in the cavity, more power is reflected back towards the amplifier, the temperature of the cavity decreases and the feedback loop is stable. To make sure it stays stable, even with changes in ambient temperature, $f_{\mathrm{res}}$ has been set around 50 kHz lower than $f_{\mathrm{sys}}$. 
%Forced air cooling was utilized to lower its final temperature.
The reflected power $P_{\mathrm{refl}}$ indicates how far the emitting cavity is off tune, and is minimized for $f_{\mathrm{res}} = f_{\mathrm{sys}}$. The relation is given in Eq.~\ref{equ:cavReflFreq},
\begin{align}
\label{equ:cavReflFreq}
\frac{P_{\mathrm{refl}}}{P_{\mathrm{inc}}} = \left|\Gamma\right|^2 = \frac{f_{\mathrm{n}}^2}{4 + f_{\mathrm{n}}^2}
\quad
f_{\mathrm{n}} = Q_{\mathrm{em}} \left(\frac{f_{\mathrm{sys}}}{f_{\mathrm{res}}} - \frac{f_{\mathrm{res}}}{f_{\mathrm{sys}}}\right)
\end{align}
where $P_{\mathrm{inc}}$ is the constant ($\pm 1 \%$) and known incident RF power at the frequency $f_{\mathrm{sys}}$. Critical coupling has been assumed, with coupling losses and impedance mismatch not considered. $P_{\mathrm{refl}}$ is measured on a directional coupler, placed between power amplifier and cavity.
%The reflected RF signal is converted to a DC voltage proportional to $P_{\mathrm{refl}}$ by a detector diode and recorded over the entire measurement time with an oscilloscope. The time trace is saved for future reference.

\begin{figure}[b]
\vspace{-0.5cm}
\begin{center}
	\includegraphics[width=0.5\textwidth]{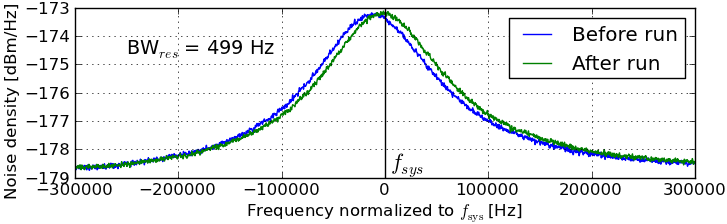}
	\caption{Measured spectral noise power density before and after the 11.5 h measurement run, indicating $f_{\mathrm{res}}$}
	\label{fig:diagTune}
\end{center}
\vspace{-0.5cm}
\end{figure}

%\subsection{Tune of the detecting cavity}
Tuning the detecting cavity once before the measurement run is sufficient, as it does not dissipate any power and changes in the laboratory's ambient temperature are small enough.
Its resonant frequency was estimated by evaluating the spectral noise power density $n_{\mathrm{o}}$ at the output of the low noise amplifier (LNA), given by Eq.~\ref{eq:thNoise}
\begin{align}
\label{eq:thNoise}
n_{\mathrm{o}} = G k_B \left[ 
T_{\mathrm{cav}} \left(1 - \left|\Gamma\right|^2\right) + 
T_{\mathrm{LNA}} + 
T_{\mathrm{LNA}}' \left|\Gamma\right|^2 
\right]
\end{align}

where $G = 44.7$~dB is the LNA's gain and $k_B$ is the boltzmann constant. Three noise temperature terms have to be considered, $T_{\mathrm{cav}} \left(1 - \left|\Gamma\right|^2\right)$ describes the noise from the detecting cavity itself. Its noise temperature is frequency dependent by the reflection coefficient $\Gamma$. The maximum occurs at $f_{\mathrm{res}}$ and is equal to ambient temperature. $T_{\mathrm{LNA}} = 32.4$~K describes the intrinsic noise temperature of the amplifier. $T_{\mathrm{LNA}}' \left|\Gamma\right|^2$ describes the noise temperature of the amplifier input, which emits a noise wave towards the cavity where it is reflected back. 
As the transmission line between amplifier and cavity is short ($\approx$ 1 cm) and as the noise temperature of the cavity walls is significantly higher than the noise wave transmitted from the amplifier's input, a good estimate of the resonant frequency can be determined from the maximum of the noise power spectrum. This has been done before and after the actual experimental run, the results are shown in Fig.~\ref{fig:diagTune} and indicate no significant (bigger than $BW_{3 \mathrm{dB}}$) drift of the emitting cavity's resonant frequency.

%\subsection{Signal processing}
The output signal of the detecting cavity is coupled out, amplified by the LNA ($G = 44.7$~dB) and then further processed by an Agilent EXA N9010A signal analyzer. The center frequency was set to $f_{\mathrm{sys}} + 4~\mathrm{Hz}$ to avoid internal spurious signals appearing at interesting parts of the spectrum. The center frequency is shifted to baseband and the complex IQ signal is digitized with 20 Hz bandwidth. The 11.5 h long time record is stored for further processing.
As the capture memory of the analyzer limits the maximum number of continuously acquired samples to $\approx 10^6$, a trade-off between recorded bandwidth and recording length had to be found. For this reason the signal analyzer will be replaced by a dedicated processing chain without this limitation for the future.

For offline data processing the spectral power of the recorded noise like signals is estimated by Welch's method implemented in a python script. The time record is read and divided into segments overlapping by $\approx 90 \%$. Each segment is multiplied by a Hann window and the complex spectra are calculated by a fast Fourier transform. Averaging these spectra trades resolution bandwidth for less variance in the noise floor. For a 11.5 h long time trace, resolution bandwidths as narrow as $BW_{\mathrm{res}} = 24~\mu$Hz can be achieved. The average noise floor is determined by $Pn = BW_{\mathrm{res}} \cdot n_o$.

The frequency error of RF-source and signal analyzer needs to be within $BW_{\mathrm{res}}$ during the experimental run, otherwise the signal power will spread out over several bins in the spectrum, degrading the signal to noise ratio. While absolute frequency drifts are unavoidable, phase-locking RF-source and signal analyzer to a common 10 MHz frequency reference allows to achieve a good relative frequency stability. This has been explained and successfully demonstrated down to $BW_{\mathrm{res}} = 10~\mu$Hz in \cite{src:narrowband}.

Immediately before the actual measurement, a test run was conducted where the shielding box was left open to provoke EM leakage between the cavities. 
The resulting power spectrum contained a single peak, clearly above the noise floor, spanning only one single bin. Its absolute position on the frequency axis was offset by $\approx 3$~mHz due to the finite resolution of the RF source. As no parameters were changed after the test run and only the shielding box was closed, the WISP related signal is expected to appear at the same position in the spectrum.
\begin{figure}[t]
\begin{center}
	\includegraphics[width=0.48\textwidth]{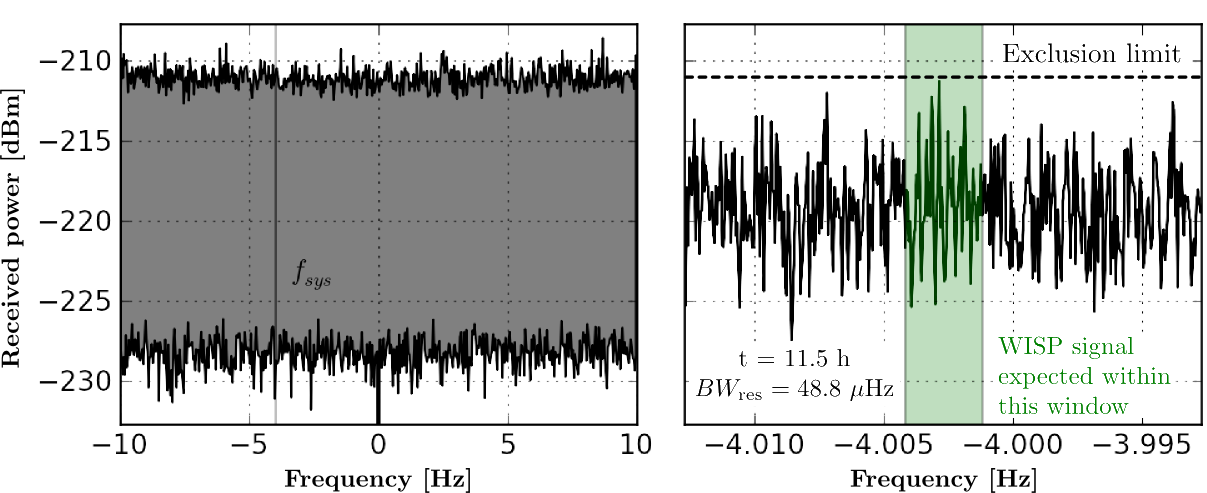}
	\caption{Spectral noise power from the detecting cavity. The left diagram shows the min. and max. peaks of the recorded span, while the right diagram is zoomed}
	\label{fig:spectra}
\end{center}
\vspace{-0.75 cm}
\end{figure} 
The peaks within a window of $\pm 1.5$~mHz around the expected signal do not exceed the peaks in other parts of the spectrum, as shown in Fig.~\ref{fig:spectra}. Therefore an exclusion result is produced by setting the minimum detectable power ($P_{\mathrm{det}} = -211$~dBm) to the maximum peak within the frequency window.

\section{Conclusion}
No paraphotons were observed in the first measurement-run of the microwave WISP search at CERN, improving the existing exclusion limits. Several technical challenges, like $> 300$~dB EM shielding between the cavities, keeping them on tune during the 11.5 h measurement run and filtering the signal with a bandwidth of $BW_{\mathrm{res}} = 24~\mu$Hz, had to be overcome.
\\
{\small
We are grateful for the practical hints and assistance from M.~Gasior and M.~Thumm.
Thanks to R.~Jones, E.~Jensen and the BE department management for support. 
}

\end{document}